\documentclass[superscriptaddress,reprint,nofootinbib,preprintnumbers,amsmath,amssymb,aps]{revtex4-2}
\usepackage{amsmath, mathrsfs, latexsym, amssymb, color, graphicx, comment, hyperref, scrhack}
\usepackage{lmodern}
\usepackage{physics}

\usepackage{graphicx}
\usepackage{color}
\usepackage[usenames,dvipsnames]{xcolor}

\begin{document}

\title{The Origin of the Gauge Theories of Glassy Systems}

\author{L.\,D.\,Son}
\affiliation{IMET UB RAS, 620016, 101 Amundsen st., Ekaterinburg, Russia}
\affiliation{Ural Federal University, 620002, 19 Myra st., Ekaterinburg, Russia}
\author{M.\,G.\,Vasin}
\affiliation{N.\,L.\,Dukhov Research Institute of Automatics (VNIIA), 127030 Moscow, Russia}
\affiliation{Vereshchagin Institute of High Pressure Physics, Russian Academy of Sciences, 108840 Moscow, Russia}

\begin{abstract}
The analytical model of a glass-forming system is formulated within the formalism analogous to gauge theory constructions in quantum field theory. This work explores the scope of the proposed approach and investigates the equilibrium behavior of the model under the mean-field approximation. The analysis reveals three possible equilibrium scenarios, only one of which exhibits strong glass-forming ability. Additionally, an upper limit for the initial temperature is identified, beyond which quenching into the glassy state becomes impossible. 
\end{abstract}

\flushbottom
\maketitle
\thispagestyle{empty}

\section{Introduction}

The glassy state remains a challenging area of research across multiple branches of condensed matter physics, spanning fields from soft matter to metallic glasses and disordered systems. This enduring scientific interest is aroused by both fundamental questions of non-equilibrium thermodynamics and practical applications in materials science. Despite extensive experimental and theoretical studies spanning decades, the fundamental microscopic mechanisms responsible for this unique state of matter are still not fully understood and continue to be vigorously debated in the scientific community.

The central challenge lies in the complex, non-equilibrium nature of glass formation, where a supercooled liquid undergoes a dramatic slowdown in dynamics without exhibiting a clear thermodynamic phase transition. While many competing theories, including the random first-order phase transition theory~\cite{PhysRevA.40.1045} and dynamical facilitation approach~\cite{Speck_2019}, have been proposed to explain the glass transition, none have yet provided a complete picture that reconciles all experimental observations. This theoretical uncertainty is further complicated by the diversity of glass-forming systems, from molecular and polymeric glasses to colloidal and metallic glasses, each exhibiting subtly different behavior while sharing common glassy characteristics.

Current research efforts focus on understanding the microscopic origins of dynamical heterogeneities, the nature of the elusive ``ideal glass state'', and the surprising universalities observed across different glass-forming systems. Advanced experimental techniques, such as ultrafast spectroscopy and high-resolution microscopy, combined with sophisticated computational approaches, are providing new insights into these long-standing questions. However, the glass transition problem remains one of the most intriguing unsolved challenges in modern condensed matter physics.

This article is devoted to one of the possible theoretical approaches to the analytical description of the glass transition, which is based on concepts of topological properties of structural distortions in locally ordered systems and quantum field methods of statistical physics. The theoretical description of the vitreous state of matter as a frozen system of topologically stable defects frustrating the ground structural state has a rather long history and was actively developed in the late 20th century.(see, e.g., \cite{Toulouse1977,Rivier1979,Morris1979,Nelson1983}).  According to this approach, topological defects (disclinations and dislocations) are considered the fundamental structural elements of glass \cite{Cheng,Hirata2013}, forming network-like structures \cite{Qi1991,Borodin1995}. 
Currently, this approach is again attracting researchers' attention \cite{Baggioli2021,Baggioli2022}.  In particular, it allows a better understanding of the plasticity properties of glasses \cite{Wu2023,Baggioli2023}.
Furthermore, theories of the glass transition based on it accurately reproduce the main thermodynamic and kinetic properties of glass formation \cite{Vasin-PRE-2022}. The primary advantage of this approach is its universality, allowing glass transitions in various physical systems to be described using a unified formalism.

The natural theoretical framework for describing topologically protected structural distortions (defects) is gauge field theory.
Gauge constructions originated in field theory as a way to introduce interaction into the Lagrangian of a free field. In general, the idea is as follows: Let there be a (quadratic) Lagrangian of a free field,
\begin{gather}\label{1}
  L=m\Psi\bar{\Psi}+c\partial_a \Psi \partial^a \bar{\Psi},  
\end{gather} 
which is invariant with respect to homogeneous (in space and time) transformations, $A$, of some continuous group $G$:
$L(\Psi, \partial_a \Psi)=L(A\Psi, A\partial_a \Psi),~~A\in G$.

When the interaction is turned on for the $\Psi$ field, the uniformity of space and time is violated. The interaction can be described if we require the invariance of the Lagrangian with respect to inhomogeneous transformations. In this case, the derivatives in (\ref{1}) are replaced by ``covariant'' ones, $\partial_a \rightarrow D_a=\partial_a +F_a^i\gamma_i$:
\begin{equation} \label{2}
~~L\rightarrow L=m\Psi\bar{\Psi}+cD_a \Psi D^a \bar{\Psi}
+F_a^i M_{ij}^{ab}F_b^j.
\end{equation}
Here $\gamma_i$ are the generators of the group $G$, $M^{ab}_{ij}$ is the interaction between defects, and $F_a^i$ are the gauge fields with spatial index $a$ and group index $i$ describing the interaction.
Constructions of the type \ref{2} are called gauge ones. The idea that such constructions can be used to describe topological defects and disorder in locally symmetric condensed media has been around for a long time \cite{kadic}. If we assume that some kinds of topological defects relax slowly, then we arrive at the critical dynamics of related fields, one of which can be considered "frozen". The presence of such a frozen field frustrates the system, i.e. causes its ground state to degenerate repeatedly.

In  \cite{Vasin-PRE-2022,Vasin-PhysicaA-2019,Vasin2} it is shown that in the gauge theory, the presence of slowly relaxing gauge fields leads to a glass transition under certain conditions. Thus, one gets the so-called gauge theory of glass. A similar approach to glass theory was originally discussed for spin systems \cite{Hertz,parisi1994}. There is also an example of the inverse effect of glass theory on field theory \cite{Ward_2022}.

The main problem with glass gauge theories is the lack of an explicit correspondence between field configurations and states of a real system. This is the main motivation of the present article, which structure is as follows.

In section 2, we consider the general formulation of a statistical model for a three-dimensional system in which the ground state is disrupted by linear topological defects and show how it corresponds to a certain gauge theory.   
Next, in section 3, we will trace how a glass transition occurs in such a system. 
In section 4, we  consider  the equilibrium behavior of the model in the mean field approximation.  
In conclusion, we discuss the possibility of getting glass by quenching.

\section{The canonical ensemble and its Hamiltonian}

Consider a three-dimensional system in which each local element is invariant with respect to the transformations of some discrete group $A$, which is a subgroup of a continuous group $B$. 
The ground (lowest-energy) state of the system is reached when all local elements correspond to the same element of group B, and small deviations lead to an increase in energy. The Hamiltonian of such a system is invariant with respect to spatially homogeneous transformations of group B. If, in the spirit of gauge theory, an inhomogeneous action of group B is allowed, then topological defects in the form of closed lines will appear in the system. When traversing the contour around the defect line, the local elements gradually transform into an element of group A, so that when the contour is closed, the local element coincides with itself.

Thus, the canonical statistical ensemble of the system under consideration is represented as a set of configurations $\{\Gamma\}$ of topological defects against the background of the ground state. Defects can be of several types (according to the elements of group A and are denoted in (\ref{H}) by the index $\Lambda$) and cannot end inside the system (they are either closed loops or endless lines).
The general form of the Hamiltonian in this case is as follows (summation over repeated indexes is suggested).
\begin{multline}\label{H}
H\{\Gamma\}=\sum_{r,k}\varepsilon_\Lambda (\alpha_k^\Lambda(r))^2\\-\frac{1}{2}\sum_{r,r'}\alpha_i^\Lambda(r)M^{ij}_{\Lambda Y}(r-r')\alpha_j^Y(r'),
\end{multline}
  where $\alpha_k^\Lambda(r)$ is the defect density tensor:
\begin{equation}\label{Alpha}
\alpha_k^\Lambda(r)=\sum_\Gamma \delta_k(r-r_\Gamma)B^\Lambda(\Gamma),
\end{equation}
and $M^{ij}_{\Lambda Y}$ is the interaction between the defects.
In (\ref{Alpha}) $\delta_k(r-r_\Gamma)$ is the $k$-th spatial component of the delta function on the defect line, $B^\Lambda(\Gamma)$ is the topological charge of the defect, and $\varepsilon_\Lambda$ is its energy per unit length. The sum covers all defects.

The Hamiltonian (\ref{H}) is used to determine the partition function and the thermodynamic potential of the system.
\begin{equation}\label{Z}
Z=\sum_{\{\Gamma\}}\exp \left(-\frac{H\{\Gamma\}}{T}\right),~~~F=-T\ln Z
\end{equation}

An analogy to gauge theory arises in the calculation of the partition (\ref{Z}). The summation over defect configurations can be replaced by a functional integral over a set of complex scalar fields $\psi^\Lambda(r)$, one field for each defect type (see, for example, \cite{PatS}). 

Let's first consider the case where there is no interaction between defects, $M^{ij}_{\Lambda Y}(r-r')=0$.

In this case, the mentioned integral on the spatial net looks as follows:
\begin{gather}\label{INT}
Z=\int\prod_{\Lambda,r}D\psi^\Lambda\left(1+\sum_d\psi^\Lambda(r)\tilde{\psi}^\Lambda(r+d)\right)e^{-\frac{|\psi^\Lambda(r)|^2}{a_\Lambda}
},
\end{gather}
where $d$ is the vector to the nearest neighbor in the spatial network, and
\begin{equation}\label{akoeff}
a_\Lambda=\exp\left(-\frac{\varepsilon_\Lambda}{T}\right) .
\end{equation}

To make sure that the equality (\ref{INT}) is fair, let's correspond the arrow of the vector $d^\Lambda$ to each product $\psi^\Lambda(r)\tilde{\psi}^\Lambda(r+d)$. When the brackets before the exponential are released, we get the sum of the different products of such arrows, with each arrow appearing no more than once. When integrated with the weight $\exp\left(-\frac{1}{a_\Lambda}
|\psi^\Lambda(r)|^2\right)$, only such products remain of this sum, where arrows of the same type $\Lambda$ form closed loops or infinite lines. Otherwise, if there are free ends, phase integration of the corresponding complex fields will yield zero at each such point.
After integration, each link of this line receives an additional multiplier $a_\Lambda$
Thus, we get the partition (\ref{Z}) without considering the interaction between the lines. To take it into account, let's use the identity (known as Hubbard--Stratonovich transformation)
\begin{multline}\label{E81}
\exp\left( \frac{1}{2T}\sum_{r,r'}\alpha_i^\Lambda(r)M^{ij}_{\Lambda Y}(r-r')\alpha_j^Y(r')\right)\\
=\int D\varphi_i^\Lambda \exp \left(-\frac{{T}}{2}\sum_{r,r'}\varphi_i^\Lambda(r)(M^{-1})^{ij}_{\Lambda Y}(r-r')\varphi_j^Y(r')\right.\\\left.
 +\sum_{r}\varphi_i^\Lambda\alpha_i^\Lambda\right),
\end{multline}
 up to an insignificant multiplier.
Thus, the term containing the tensor $\alpha_i^\Lambda(r)$ becomes local, namely
\begin{equation}\label{E9}
 \exp\left(\sum_r\varphi_i^\Lambda
 \alpha_i^\Lambda\right)   
\end{equation}

Here, it is natural to introduce the auxiliary  field  $\xi_i^\Lambda$
\begin{equation}\label{E10}
 \exp\left(\sum_r\varphi_i^\Lambda
 \alpha_i^\Lambda\right)\rightarrow  \exp\left(\sum_r(\varphi_i^\Lambda+\xi_i^\Lambda)
 \alpha_i^\Lambda\right),  
\end{equation}
which allows one to get easily various averages of the 
of the field $\alpha_i^\Lambda$ by the derivatives with respect to $\xi_i^\Lambda$. For example,
\begin{equation}
    \langle \alpha_i^\Lambda(r)\rangle=\frac{\partial \ln Z}{\partial\xi_i^\Lambda(r)}|_{\xi_i^\Lambda\rightarrow 0}
\end{equation}
Right now, we omit this term, but it may be easily restored at any stage.

To account (\ref{E9}), one only needs to make a change in (\ref{INT}):
\begin{multline}
\sum_d \psi^\Lambda(r)\tilde{\psi}^\Lambda(r+d)\rightarrow \\\sum_d\psi^\Lambda(r)\tilde{\psi}^\Lambda(r+d)\exp
\left(d_i B^\Lambda \varphi_i^\Lambda(r)\right),
\end{multline}
where $\sum\limits_d$ is the neighborhood summation,

Let us redefine:
\begin{equation}
B^\Lambda\varphi_i^\Lambda \rightarrow \varphi_i^\Lambda,~~~ \frac{{T}\left(M^{-1}\right)^{ij}_{\Lambda Y}}{B^\Lambda B^Y}\rightarrow {T}\left(M^{-1}\right)^{ij}_{\Lambda Y}.
\end{equation}
Thus, we come to the expression (in continual form)
\begin{gather}\label{Z}
 Z=\int D\psi^\Lambda D\varphi_i^\Lambda\exp \left(-F\{ \psi^\Lambda,\varphi_i^\Lambda\}\right),   
\end{gather} 
\begin{multline}
F\{ \psi^\Lambda, \varphi_i^\Lambda\}=\sum_{r,\Lambda}\frac{1}{a_\Lambda}|\psi^\Lambda(r)|^2
\\+\frac{T}{2}\sum_{r,r'}\varphi_i^\Lambda(r)(M^{-1})^{ij}_{\Lambda Y}(r-r')\varphi_j^Y(r')\\
-\left[\sum_{r,\Lambda}\ln\left(1+\nu|\psi^\Lambda(r)|^2+
\gamma\psi^\Lambda(\partial_i+\varphi_i^\Lambda)^2\tilde{\psi}^\Lambda\right)\right] \label{F},
\end{multline}
%{\color{red}
%Maybe eliminate the $\alpha$ field altogether? Then 
%\begin{eqnarray}
%Z=\int D\psi^\Lambda D\varphi_i^\Lambda\exp \left(-F\{ \psi^\Lambda,\varphi_i^\Lambda\}\right), \nonumber\\ F\{ \psi^\Lambda, \varphi_i^\Lambda\}=\sum_{r,\Lambda}\varphi_i^{\Lambda}(r)\psi^\Lambda(r) +\frac{T}{2}\sum_{r,r'}\varphi_i^\Lambda(r)(M^{-1})^{ij}_{\Lambda Y}(r-r')\varphi_j^Y(r')-\nonumber\\ -\left[\sum_{r,\Lambda}\ln\left(1+\nu|\psi^\Lambda(r)|^2+ \gamma\psi^\Lambda(\partial_i+\varphi_i^\Lambda)^2\psi^\Lambda\right)\right] \label{F} \end{eqnarray}
%}
where, $\nu$ is the number of nearest neighbors in the spatial network. 

To get this, one has to use the following expansions:
\begin{gather}
\tilde{\psi}^\Lambda (r+d) \simeq \tilde{\psi}^\Lambda (r)+d_i\partial_i\tilde{\psi}^\Lambda(r) +\frac{1}{2}
d_id_k\partial^2_{ik}\tilde{\psi}^\Lambda(r), \nonumber\\
\exp\left(d_i\varphi_i^\Lambda\right)\simeq
1+d_i\varphi_i^\Lambda+\frac{1}{2}d_i d_k
\varphi_i^\Lambda\varphi_k^\Lambda, \nonumber
\end{gather}
and the equalities:
\begin{equation}
\sum_d d_i=0,~~~\frac{1}{2} d_i d_k=\gamma d^2\delta_{ik}
\end{equation}

In the low-energy approximation, if instead of the logarithm we can limit ourselves to its expansion up to the fourth-order terms, we have 
\begin{multline}\label{FBase}
F\{ \psi^\Lambda, \varphi_i^\Lambda\}\cong \sum_{r,\Lambda}\left(\left(\frac{1}{a_\Lambda}-\nu\right)|\psi^\Lambda|^2
\right.\\\left.-\gamma \psi\left( D_k^\Lambda\right)^2 \tilde{\psi}^\Lambda+\frac{\nu^2}{2}|\psi^\Lambda|^4\right)\\
+\frac{T}{2}\sum_{r,r'}\varphi_i^\Lambda(r)(M^{-1})^{ij}_{\Lambda Y}(r-r')\varphi_j^Y(r'),
\end{multline}
where
\begin{equation}
D_k^\Lambda=\partial_k+\varphi_k^\Lambda ,
\end{equation}
 and $\varphi^{\Lambda}_k$ field acts as the gauge field corresponding to the defect interaction.
 
The following elements should be stressed.
\begin{enumerate}
\item The analogy with gauge theories takes place only in the low-energy approximation.
\item The core of the interaction $M_{\Lambda Y}^{ij}(r-r')$ is determined by the Green function of the ground state, which is a separate task which, in turn, requires separate experimental and theoretical efforts.
\item There is no one-to-one correspondence between the individual configurations of fields and defects; the correspondence is only in the language of averages. For example, a nonzero average value of the $\psi$ field means the appearance of infinite defects, and a nonzero average vector $\vec{\varphi}$ means their orientation ordering.
\item $\langle \vec{\varphi}\rangle \neq 0 $ means that certain kinds of defects are uncompensated, so that the ground state cannot be restored uniquely, i.e., frustrations take place.
\item The $\psi$ field has an obvious critical point
\begin{equation}\label{TC}
T_c=\varepsilon/\ln \nu,
\end{equation}
 since the coefficient $\left(1/a_\Lambda-\nu\right)$ passes through zero.
\end{enumerate}
Note that, according to (12), the statistical description of the system corresponds to a statistical theory with a non-equilibrium thermodynamic potential (or, in another terminology, an effective Hamiltonian)
\begin{equation}
\mathcal{F}\{\psi^\Lambda,\varphi_i^\Lambda\}=TF\{\psi^\Lambda,\varphi_i^\Lambda\}
\end{equation}
In the next section, we show how the glass transition occurs in the low-energy approximation. To do so, we use critical dynamics methods in the spirit of works \cite{Vasin-PhysicaA-2019} -- \cite{Vasin-PRE-2022}

\section{Kinetics in the Low-Energy Approximation}

The system's kinetics is described by the following  stochastic Langevin  equations:
\begin{eqnarray}
\Gamma_{\psi}\frac{\partial \psi^\Lambda}{\partial t}=-\frac{\delta \mathcal{F}\{ \psi^\Lambda, \varphi_i^\Lambda\}}{\delta \psi^\Lambda}+\xi_{\psi}^\Lambda, \\
\Gamma_{\varphi}\frac{\partial \varphi^\Lambda}{\partial t}=-\frac{\delta \mathcal{F}\{ \psi^\Lambda, \varphi_i^\Lambda\}}{\delta \varphi^\Lambda}+\xi_{\varphi}^\Lambda,
\end{eqnarray}
where $\langle\xi_{\psi}^\Lambda  \xi_{\psi}^\Lambda\rangle_{{\bf r},\,t}=\Gamma_{\psi}T\delta_{\bf r}\delta_t$, $\langle\xi_{\varphi}^\Lambda\xi_{\varphi}^\Lambda \rangle=\Gamma_{\varphi}T\delta_{\bf r}\delta_t$, $\Gamma_\psi$ and $\Gamma_{\varphi}$ are the kinetic parameters.
The statistical properties of this system can be expressed with the use of the following generating functional:
%\newpage
\begin{multline}
W=
\int D\psi^\Lambda D\varphi_i^\Lambda\prod\limits_{t=-\infty}^{\infty} \delta \left(\Gamma_{\psi}\partial_t \psi^\Lambda +\frac{\delta \mathcal{F}}{\delta \psi^\Lambda}-\xi_{\psi}^\Lambda\right)\\\times\delta \left(\Gamma_{\varphi}\partial_t \varphi^\Lambda+\frac{\delta \mathcal{F}}{\delta \varphi^\Lambda}-\xi_{\varphi}^\Lambda\right)\\=
\int D\psi^\Lambda D\varphi_i^\Lambda D{\psi'}^\Lambda D{\varphi'}_i^\Lambda e^{ i\int\limits_{-\infty}^{\infty}\mathrm{d}t\,\mathcal{L}'\{ \psi^\Lambda,{\psi'}^\Lambda, \varphi_i^\Lambda,{\phi'}_i^\Lambda\}},
\end{multline}
where
\begin{multline}
\mathcal{L}'\{ \psi^\Lambda,{\psi'}^\Lambda, \varphi_i^\Lambda,{\phi'}_i^\Lambda,\}= {\psi'}^\Lambda\Gamma_{\psi}\partial_t \psi^\Lambda \\+{\psi'}^\Lambda\frac{\delta \mathcal{F}\{ \psi^\Lambda, \varphi_i^\Lambda\}}{\delta \psi^\Lambda}-{\psi'}^\Lambda\xi_{\psi}^\Lambda +
{\varphi'}_i^\Lambda\Gamma_{\varphi}\partial_t\varphi_i^\Lambda\\+{\varphi'}_i^\Lambda\frac{\delta \mathcal{F}\{ \psi^\Lambda \varphi_i^\Lambda\}}{\delta \varphi_i^\Lambda}-{\varphi'}_i^\Lambda\xi_{\varphi}^\Lambda
\end{multline}
is the effective Lagrangian.
After Wick rotation and averaging over $\xi_\psi$ and $\xi_{\varphi}$   
the effective Lagrangian
looks as follows:
\begin{multline}
\mathcal{L}\{ \psi^\Lambda,{\psi'}^\Lambda, \varphi_i^\Lambda,{\varphi'}_i^\Lambda,\}= {\psi'}^\Lambda\Gamma_{\psi}\partial_t \psi^\Lambda \\+{\psi'}^\Lambda\frac{\delta \mathcal{F}\{ \psi^\Lambda, \varphi_i^\Lambda\}}{\delta \psi^\Lambda}+\Gamma_{\psi}T{\psi'}^\Lambda{\psi'}^\Lambda +
{\varphi'}_i^\Lambda\Gamma_{\varphi}\partial_t\varphi_i^\Lambda\\+{\varphi'}_i^\Lambda\frac{\delta \mathcal{F}\{ \psi^\Lambda \varphi_i^\Lambda\}}{\delta \varphi_i^\Lambda}+\Gamma_{\varphi}T{\varphi'}_i^\Lambda{\varphi'}_i^\Lambda.
\end{multline}

Let us restrict ourselves to a single type of topological defect, so that the index $\Lambda$ is not necessary. In that case, (\ref{FBase}) looks as  
\begin{multline}
\mathcal{F}\{ \psi, \varphi_i \} \cong T\sum_{r} \left(u|\psi|^2
-\gamma \psi(\partial_k+\varphi_k)^2\tilde{\psi}\right.\\+\left.\frac{\nu^2}{2}|\psi|^4\right)
+\frac{T^2}{2}\sum_{r,r'}\varphi_i(r)(M^{-1})^{ij}(r-r')\varphi_j(r'), \label{E36}
\end{multline}
where
\begin{equation}
u=\exp\left(\frac{\varepsilon}{T}\right) -\nu .
\end{equation}

For the ``gauge'' fields interaction, let us choose the simplest reasonable form:
\begin{equation}
(M^{-1})^{ij}(r-r')= \delta_{ij}\delta(r-r')\left(m-c\nabla^2\right),
\end{equation}
which provides a preferable parallel orientation of neighboring defects and decay of interaction at distance $\sim \sqrt{c/m}$.

Thus the effective Lagrangian looks as follows:
\begin{multline}
\mathcal{L}\{ \psi,{\psi'}, \varphi_i,{\varphi'}_i,\}
= {\psi'}\Gamma_{\psi}\partial_t \psi \\+{\psi'} T\sum_{r} \left(2u \psi
-\gamma (\nabla+\vec\varphi+B\vec\xi)^2\tilde{\psi}+2\nu^2|\psi|^3\right) \\+ \vec{\varphi}'\Gamma_{\varphi}\partial_t\vec{\varphi}-\vec{\varphi}' T\sum_{r}\left(
2\gamma \psi(\nabla+\vec{\varphi}+B\vec{\xi})\tilde{\psi}
\right.\\\left.
-T^2(m-c\nabla^2)\vec{\varphi}\right)+
\Gamma_{\psi}T{\psi'}{\psi'} 
+\Gamma_{\varphi}T\vec{\varphi}'\vec{\varphi}'.\label{L}
\end{multline}
The critical kinetics of systems like this, in which the evolution of the observed order parameter depends on the kinetics of its gauge field, was described previously in \cite{Vasin-PhysicaA-2019,Vasin-PRE-2022,Vasin1}. Indeed, one can show that the relaxation of the considered system has a slow, non-Arrhenius character.  

In the $({\bf k},\,t)$ - representation, in the long-wave limit (${\bf k}\to 0$),  the (\ref{L}) can be rewritten as follows:
\begin{multline}\label{L1}
\mathcal{L}\{ \psi,{\psi'}, \varphi_i,{\varphi'}_i,\}
= {\psi'}(\Gamma_{\psi}\partial_t+T(2u -\gamma \vec\varphi^2)){\psi}\\+2\nu^2T{\psi'} |\psi|^3 +
\vec{\varphi}'(\Gamma_{\varphi}\partial_t+ T(
Tm-2\gamma \psi^2))\vec{\varphi}\\
+\Gamma_{\psi}T{\psi'}{\psi'} 
+\Gamma_{\varphi}T\vec{\varphi}'\vec{\varphi}'.
\end{multline}
From this expression, one can see that at some non-zero average value of
\begin{equation} \label{Tg1} 
|\langle\vec\varphi\rangle|=\sqrt{2u/\gamma}, 
\end{equation}
the $\psi$ field relaxation time becomes infinitely large.

To estimate the relaxation time, note
that from (\ref{E81}) one can show (\cite{Vasin-PRE-2022}) that 
$\langle\vec\alpha'\vec\alpha\rangle\propto\exp(\mbox{const}\langle\vec\varphi'\vec\varphi\rangle)$. In the same time in momentum space,
\begin{equation}\label{psi-psi}
   \langle\vec\varphi'\vec\varphi\rangle_{{\bf k},\omega}\propto
   \frac{1}{{\bf k}^2+i\Gamma\omega +m},
\end{equation}
where $m\propto T-T_g$, and the glass transition temperature is defined by (\ref{Tg1}):
\begin{equation}
  T_g=\frac{\varepsilon}{\ln(\nu+\frac{\gamma}{2}\langle \vec{\varphi}\rangle^2)}.  
\end{equation}
  In the limit ${\bf k}\to0,\omega\to 0$ the (\ref{psi-psi}) gives the following  temperature dependence of the relaxation time:
\begin{eqnarray}\label{E38}
\tau\propto \langle\vec\alpha'\vec\alpha\rangle_{k= 0,\omega=0}\propto \exp\left(\frac{\mbox{const}}{T-T_g}\right),
\end{eqnarray}
that corresponds to the non-Arrhenius slow relaxation near the glass transition temperature (Vogel--Fulcher--Tammann law). 
Let us underline that glass transition requires effective frustration of the system by some starting nonzero mean value of the gauge field $\langle\vec{\varphi}\rangle$, which corresponds to the presence of uncompensated topological defects. To provide this condition, it is necessary that the system has an equilibrium state with a non-zero $\langle\vec{\varphi}\rangle$ that can be supercooled. Thus, we need to investigate the behavior of the system outside the low-energy approximation, which is done in the next section.

\section{Mean Field Approximation}

In the previous section, we studied the system in the low-energy approximation, where we can limit ourselves to the first terms in the logarithmic expansion. However, it is unclear whether there is a way to create a non-zero average density of uncompensated defects in the system. To find out, we need to study the equilibrium behavior of the system without assuming that the fields are small. This can be done using the mean-field approximation. We can use the following expression for the thermodynamic potential as a function of the mean fields $\psi$ and $\vec{\varphi}$:
\begin{equation}\label{FLandau}
F=a\psi^2+b\varphi^2-\ln\left(1+\psi^2+\psi^2\varphi^2\right).
\end{equation}
This is functional (\ref{FBase}) written for the single type of defects. 

Here, we suggested homogeneous fields $\vec{\varphi},\psi$ and used the substitution
\begin{equation}
\psi\sqrt{\nu}\to\psi,~~\varphi_i\sqrt{\frac{\gamma}{\nu}}\to\varphi_i
\end{equation}
to avoid phenomenological constants under the logarithm, and notations
\begin{equation}
a=\frac{1}{\nu}\exp\left(\frac{\varepsilon}{T}\right),~~~b=\frac{T}{M},
\end{equation}
with $M$ defined by
\begin{equation}
\frac{\gamma}{2\nu}\int  M_{ik}(r-r')dr'=-M\delta_{ik}.
\end{equation}
Expression (\ref{FLandau}) may be treated as a non-equilibrium thermodynamic potential in the Landau theory. We examine the phases of the model, for which we determine the extremes (minima) of the potential (\ref{FLandau}) as a function of ($\psi, \vec{\varphi}$). There are 4 parameters in the model - $\nu,\gamma,\varepsilon, M$. The first two are geometric, the second two are energetic.  
The mean-field equations are:
\begin{eqnarray}
a\psi&=&\frac{\psi(1+\varphi^2)}{1+\psi^2(1 +\varphi^2)}, \label{PSI}\\
b\varphi&=&\frac{\varphi}{\varphi^2+(1+\psi^2)/\psi^2}.\label{PHI}
\end{eqnarray}
Note, that potential (\ref{FLandau}) includes only squares of absolute values of $\psi,\vec{\varphi}$, so that (\ref{PHI},\ref{PSI}) should be understood as equations on these absolute values. These equations have a single solution $\psi=0,~\varphi=0$ at low temperatures that corresponds to the ground state of the system.
This solution becomes unstable at the temperature (\ref{TC}). Above this temperature, a non-zero $\psi$, i.e. a finite density of defect lines appears, and the system goes into a state with $\psi\neq0$.

 From (\ref{PHI},\ref{PSI}) closed equation on $\varphi$ may be derived:
 \begin{equation}\label{phi^4}
     \varphi^4-\left(\frac{1}{b}-2\right)\varphi^2 +\frac{a+b-1}{b}=0.
 \end{equation}

 Its analysis leads to the following three scenarios of the system's behavior:
 
 1) At low $M$, equation (44) has no solutions at any temperature, and the equilibrium value $\varphi$ remains zero (frustrations do not occur). There are two phases in the system – the ground state ($\psi=0,\varphi=0$) and the state with a finite density of randomly located defects ( $\psi\neq 0,\varphi=0$). They are separated by a transition of the second kind at temperature (\ref{TC}). In this case, it will not be possible to make glass in the system – there is no defective state that would have stability below (\ref{TC}), i.e., which could be quenched.

 2) There exists a single positive solution of (\ref{phi^4}).  This means that there is a region where the free term in (\ref{phi^4}) is less than zero:
 \begin{equation}
     a+b-1=\frac{\exp\{\varepsilon/T\}}{\nu}+\frac{T}{M}-1<0.
 \end{equation}
This inequality defines the temperature range  $(T_-,T_+ )$, at the boundaries of which the single nonzero and positive solution for $\varphi^2$ appears and disappears. Also, the second coefficient  in (44) should be positive:
\begin{equation}
    \frac{M}{T}<2.
\end{equation}
Then, the appearance and disappearance occur at $\varphi^2=0$. At the same time, the  point $\varphi^2=0$ itself (let us remind that there is always an extrema in it) turns into a maximum, and a non-zero solution corresponds to a minimum.
Transitions at points $T_-,T_+$ are then transitions of the second kind. The scenario of behavior with increasing temperature is as follows: the basic state $\to$ the appearance of defects at $T_c~\to$ ordering of defects at $T_-~\to$ disordering of defects at $T_+$. Quenching into a glassy state is possible from the interval $(T_-,T_+ )$, but is rather difficult: it requires a high cooling speed because it is necessary to hold  nonzero $\varphi$ below $T_-$, where it is not divided from zero by any energy barriers.

3) There exist two positive solutions of (\ref{phi^4}). To provide this, the second coefficient  in (44) should be negative:
\begin{equation}
    \frac{M}{T}>2,
\end{equation}
the free term and the discriminant should be positive:
\begin{eqnarray}
     a+b-1=\frac{\exp\{\varepsilon/T\}}{\nu}+\frac{T}{M}-1>0,\\
     \frac{1}{b}-4a=\frac{M}{T}-\frac{4\exp\{\varepsilon/T\}}{\nu}>0
\end{eqnarray}
The last is possible if
\begin{equation}
    \frac{M\nu}{4\varepsilon}>e.
\end{equation}
Thus, the two positive solutions exist in the temperature interval
\begin{equation}\label{interval}
    T_{down}=\frac{\varepsilon}{x_1}<T<\frac{\varepsilon}{x_2}=T_{up}=T_+,
\end{equation}
where $x_1, x_2$ - upper and lower roots of the equation
\begin{equation}
    \frac{M\nu}{4\varepsilon}x=e^x.
\end{equation}
This interval may be wide enough.  In this interval, the state with uncompensated defects $\psi\neq0,\varphi\neq0$ is stable (or metastable) and corresponds to the minima of thermodynamic potential.  $T_{down}$ is the low spinodal temperature of the $\psi\neq0,\varphi\neq0$ state. The transition between  $\psi\neq0,\varphi=0$ and $\psi\neq0,\varphi\neq0$ is the first order phase transition. The temperature $T_-$ is the spinodal temperature of the phase $\psi\neq0,\varphi=0$. Between $T_{down}$ and $T_-$
the state $\psi\neq0,\varphi\neq0$ is
separated by energy barrier from the states without frustrations ($\psi=0,\varphi=0$ and $\psi\neq0,\varphi=0$). Equilibrium minima and barrier maxima correspond to the upper and lower roots of (\ref{phi^4}), respectively. At $T_-$ the last goes to zero, so that the phase $\psi\neq0,\varphi=0$ becomes unstable. The temperature $T_up$, where the discriminant of (\ref{phi^4}) is zero, coincides with $T_+$. At this temperature, discriminant, second and free term of (\ref{phi^4}) become zeros, so that at this temperature reverse second order phase transition from $\psi\neq0,\varphi\neq0$ to $\psi\neq0,\varphi=0$ takes place. 

Note, that the equilibrium behavior may imply a direct first order transition from the ground state to the $\psi\neq0,\varphi\neq0$. In that case, $T_c$ is the spinodal of the ground state, and $\psi\neq0,\varphi=0$ is an intermediate metastable phase.

The third scenario of the equilibrium behavior provides the best conditions for the glass formation via quenching.

\section{Conclusion}

In this work, we model three-dimensional systems where a well-defined ground state is disrupted by topological defects. We analyze the system's behavior both within the mean-field approximation and in the low-energy limit, focusing on the simplest scenario involving a single type of topological defect. This case may also describe systems where the defect dynamics can be effectively captured by a representative ``mean type'' of defect.

In this work, we study general behavioral patterns in models where topological defects frustrate a well-defined ground state. Thus,
1) We show that the partition function of the model may be written as a functional integral over two fields with some effective Hamiltonian (\ref{Z}). The integration over the complex scalar field generates a summation over configurations of topological defects, and the integration over the vector field accounts for their interaction.
2) In the low-energy limit, when one can use a power expansion in the effective Hamiltonian, the theory looks like a gauge field theory (\ref{FBase}). By analyzing the associated Langevin stochastic equations, we show that a system initially frustrated by uncompensated defects can be quenched to a glassy state (see section 3). 
3) To get initial frustration, $\langle\vec{\varphi}\rangle \neq 0$, it is necessary that the phase $\psi\neq0,\varphi\neq0$ be in equilibrium in some temperature interval and be separated from the ground state by an energy barrier, i.e. between these states must exist a first order phase transition. Thus, the temperature interval (\ref{interval}) should be wide enough,
 \begin{equation}
     \frac{M\nu}{4\varepsilon}\gg e.
 \end{equation}
In other words, the defect interaction energy should be greater than the core energy of the defect, and the third scenario of the mean field behavior should be realized; 4) It should be stressed that quenching to the glassy state is difficult if in the initial state there is not a stable average value of the gauge field, $\langle{\varphi}\rangle \neq 0$, in particular if the quenching starts from temperatures above $T_{up}$. This explains why metallic glasses cannot be obtained if the initial melt is overheated.

%{\bf Acknowledgments}

%The work was supported by RSF (project )

%\bibliographystyle{elsarticle-num} 
%\bibliography{ssylki_gauge}

\end{document}